\documentclass[prb,twocolumn,nobibnotes,groupedaddress]{revtex4}

\usepackage{algorithm,algorithmic}
\usepackage{enumerate}
\usepackage{listings}
\usepackage{color}
\usepackage{float}
\usepackage{amssymb}
\usepackage{amsmath}
\usepackage{tikz}
\usetikzlibrary{quantikz}

\usepackage{graphicx}

\definecolor{red}{rgb}{1,0.,0}

\begin{document}

\title{Improving the accuracy and efficiency of quantum connected moments expansions}
\date{\today}
\author{Daniel Claudino,$^{1,2}$ Bo Peng,$^{3}$ Nicholas P. Bauman,$^{3}$ Karol Kowalski$^{3}$ and Travis S. Humble$^{1,4}$}
\affiliation{$^{1}$Quantum Computing Institute,\ Oak\ Ridge\ National\ Laboratory,\ Oak\ Ridge,\ TN,\ 37831,\ USA \\
$^{2}$Computer Science and Mathematics,\ Oak\ Ridge\ National\ Laboratory,\ Oak\ Ridge,\ TN,\ 37831,\ USA \\
$^{3}$Physical Sciences Division, Pacific Northwest National Laboratory, Richland, WA 99352, USA\\
$^{4}$Computational Sciences and Engineering,\ Oak\ Ridge\ National\ Laboratory,\ Oak\ Ridge,\ TN,\ 37831,\ USA}

\begin{abstract}

The still-maturing noisy intermediate-scale quantum (NISQ) technology faces strict limitations on the algorithms that can be implemented efficiently. In the realm of quantum chemistry, the variational quantum eigensolver (VQE) algorithm has become ubiquitous, with many variations. Common to these approaches is the use of the functional form of the ansatz as a degree of freedom, whose parameters are determined variationally in a feedback loop between the quantum processor and its conventional counterpart. 
Alternatively, a promising new avenue has been unraveled by the quantum variants of techniques grounded on expansions of the moments of the Hamiltonian, among which two stand out: the connected moments expansion (CMX) [Phys. Rev. Lett. \textbf{58}, 53 (1987)] and the Peeters--Devreese--Soldatov (PDS) functional [J. Phys. A \textbf{17}, 625 (1984); Int. J. Mod. Phys. B \textbf{9}, 2899], the latter based on the standard moments $\langle \hat{H}^k \rangle$. Contrasting with VQE-based methods and provided the quantum circuit prepares a state with non-vanishing overlap with the true ground state, CMX often converges to the ground state energy, while PDS is guaranteed to converge by virtue of being variational.
However, for a finite CMX/PDS order, the circuit may significantly impact the energy accuracy. Here we use the ADAPT-VQE algorithm to test shallow circuit construction strategies that are not expected to impede their implementation in the present quantum hardware while granting sizable accuracy improvement in the computed ground state energies. We also show that we can take advantage of the fact that the terms in the connected moments are highly recurring in different powers, incurring a sizable reduction in the number of necessary measurements. By coupling this measurement caching with a threshold that determines whether a given term is to be measured based on its associated scalar coefficient, we observe a further reduction in the number of circuit implementations while allowing for tunable accuracy.

\end{abstract}

\thanks{This manuscript has been authored by UT-Battelle, LLC, under Contract No.~DE-AC0500OR22725 with the U.S.~Department of Energy. The United States Government retains and the publisher, by accepting the article for publication, acknowledges that the United States Government retains a non-exclusive, paid-up, irrevocable, world-wide license to publish or reproduce the published form of this manuscript, or allow others to do so, for the United States Government purposes. The Department of Energy will provide public access to these results of federally sponsored research in accordance with the DOE Public Access Plan.}

\maketitle
\section{Introduction}

While fault-tolerant quantum computation is still not on the horizon, the attempt to take advantage of the current noisy intermediate-scale quantum (NISQ)\cite{Preskill2018quantumcomputingin} devices has spurred a whole slew of algorithms that cleverly aim at circumventing the limitations of the present technology. Successful algorithms in the NISQ era have shown a careful design in favor of a concerted operation of classical and quantum processors, where the latter is in charge of tasks the former is unable to efficiently perform, i.e., state preparation and measurement. One such algorithms is the variational quantum eigensolver (VQE),\cite{vqe} whose foundation on the variational principle delivers an upper bound to the lowest eigenvalue of an observable, more commonly the ground state energy of a given Hamiltonian. The variational principle is in close connection with methods in quantum chemistry, so it should come as no surprise the readily application of VQE to chemical systems.

As dictated by the variational principle, the energy computed by VQE can be made more accurate by the inclusion of more variational parameters in the trial wave function/circuit ansatz. On the other hand, as the dimensionality of the corresponding Hilbert space increases, the number of such parameters necessary to satisfactorily cover the relevant part of the Hilbert space can grow out of proportion, eventually becoming unfeasible for implementation in the current quantum hardware. An alternative approach is to re-engineer how the ground state energy is estimated for some trial wave function. This is at the heart of the different expansions in terms of the moments of the Hamiltonian\cite{horn1984t} and their recent rise in the quantum computation of approximate ground-state energies of chemical systems.

For a given such expansion of order $K$, one needs to compute all moments up to $2K-1$ ($\langle H^n \rangle$), leading to a measurement schedule that can become overwhelming if there is a need to go to high orders in the expansion. Even though there can be several flavors of quantum expansions of this type, they all share the fact that, as long as the quantum state from which the moments of the Hamiltonian has non-vanishing overlap with the true ground state, they are prone to recover the ground state energy. However, it does not mean that all expansions display identical convergence, as can be seen in Ref. \citenum{kowalski2020cmx}. Moreover, the convergence can be accelerated if the measurements are based on a state with increasing overlap with the targeted ground state. 

In previous papers,\cite{kowalski2020cmx,peng2021variational} it was shown that the Peeters--Devreese--Soldatov energy functional, PDS,\cite{peeters1984upper, soldatov1995generalized} displays remarkable convergence properties, even with moments computed from a Hartree--Fock state. In order to strike a balance between circuit depth and the number of measurements, here we investigate some potential avenues for generating shallow circuits that are still effective in increasing the overlap with the exact ground state, which is evidenced by improved energy accuracy. For this purpose, we find that the iterative nature of the Adaptive Derivative-Assembled Pseudo-Trotter VQE, ADAPT-VQE,\cite{adapt} provides a natural and intuitive framework for generating such circuits, which, in combination with the low-order PDS(2) and CMX(2) expansions, signals remarkable improvement, without incurring significant resource overhead. Another target this work aims at is the reduction of the number of necessary measurements, each of which requires a circuit implementation. We recognize that higher moment expansion orders involve powers of the Hamiltonian that are increasingly composed of terms found in lower-order moments, thus can be cached upon measurement and accessed once the corresponding term is later encountered. The efficiency gains granted by this caching can be further extended by identifying that some terms have a negligible contribution and can thus be left out of the measurement schedule.

\section{Theory}
\label{sec:theory}

In this section we provide only salient features of the underlying methodologies employed herein, while pointing to the specific bibliography where pertinent. 
\subsection{CMX Expansion}
The Horn--Weinstein (HW) theorem \cite{horn1984t}  or  ``$t$-expansion''  enables one to express the function $E(t)$  defined as 
\begin{equation}
    E(t)=\frac{\langle\Phi|\hat{H}e^{-t \hat{H}}|\Phi\rangle}
    {\langle\Phi|e^{-t \hat{H}}|\Phi\rangle} 
    \label{hw1}
\end{equation}
in terms of the so-called connected moments $I_k$ 
\begin{equation}
E(t) = \sum_{k=0}^{\infty} \frac{(-t)^k}{k!} I_{k+1}
    \;,
    \label{hw2}
\end{equation}
where 
\begin{equation}
    I_k=\langle\Phi|\hat{H}^k|\Phi\rangle - \sum_{i=0}^{k-2}
    {k-1 \choose i} I_{i+1} \langle\Phi|\hat{H}^{k-i-1}|\Phi\rangle.
    \label{hw3}
\end{equation}
It is also well-known that if the trial wave function $|\Phi\rangle$ has non-zero overlap with exact ground-state  eigenstate  $|\Psi\rangle$ of the many-body Hamiltonian $H$,  then  $E(t)$ converges in the $t \rightarrow \infty $ limit  to the 
the exact ground-state energy $E_0$. 
The initial applications of $t$-expansion to describe energies of many-body systems were performed using various types of Pad\'e approximants. 
A more practical approach, that directly addresses the algebraic form of the $E(t)$ in the $t \rightarrow \infty$ limit, was introduced by Cioslowski in Ref.\citenum{cioslowski1987connected} who recast $E(t)$
in the form of decreasing exponentials
\begin{equation}
    E(t) = E_0 + \sum_{j=1}^{\infty} A_j e^{-b_j t} \;, (b_j > 0) \;,
    \label{hw4}
\end{equation}
which lead to the connected moments expansion for $E_0$
\begin{eqnarray}
    E_0=I_1-\frac{S_{2,1}^2}{S_{3,1}}(1+\frac{S_{2,2}^2}{S_{2,1}^2S_{3,2}}\ldots (1+\frac{S_{2,m}^2}{S_{2,m-1}^2S_{3,m}})\ldots), \nonumber\\
 \label{hw5}
\end{eqnarray}
where $S_{k,1}=I_k~~(k=2,3,\ldots)$ and $S_{k,i+1}=S_{k,1}S_{k+2,i}-S_{k+1,i}^2$.
The truncation of the CMX series (\ref{hw5}) provides  CMX($K$) approximations. For example, the CMX(2) and CMX(3) energies are defined as follows,
\begin{eqnarray}
    E_0^{\rm CMX(2)}&=& I_1 - \frac{I_2^2}{I_3} \;,\;\;\label{cmx2} \\
    E_0^{\rm CMX(3)}&=& I_1 - \frac{I_2^2}{I_3}
    -\frac{1}{I_3} \frac{(I_2I_4-I_3^2)^2}{I_5I_3-I_4^2}\;.\label{cmx3}
\end{eqnarray}
The``$t$-expansion'' and CMX approximations were tested and verified on benchmark many-body systems encountered in quantum chemistry and condensed matter physics.
\cite{knowles1987validity,stubbins1988methods,perez1988t,yoshida1988connected,ullah1995removal,mancini1994analytic,mancini1997numerical,piecuch2003exactness,zhuravlev2016cumulant, Vallury2020quantumcomputed}

\subsection{PDS energy functional}
An interesting  formulation to calculate ground- and excited-state energies, and based on the utilization of standard/canonical moments $\langle\Phi|H^i|\Phi\rangle$, has  been proposed by
Peeters and Devreese \cite{peeters1984upper} (further analyzed by
Soldatov
\cite{soldatov1995generalized}) where the upper bound of the ground-state energy in $K$-th order approximation (denoted as PDS($K$)) is calculated as a root of the polynomial $P_K(x)$
\begin{equation}
    P_K(x) = \sum_{i=0}^K a_i x^{K-i} \;, 
    \label{hw6}
\end{equation}
where
$a_0=1$
and
coefficients $a_i$'s for $1\le i \le K$
(forming vector ${\bf a}$) are obtained by solving linear equations
\begin{equation}
    {\bf M}{\bf a}=-{\bf b}
\end{equation}
with matrix elements  ${\bf M}_{ij}=\langle\Phi|\hat{H}^{2K-(i+j)}|\Phi\rangle$ and vector component 
$b_i=\langle\Phi|\hat{H}^{2K-i}|\Phi\rangle$.
The exact ground-state energy of a system can be calculated as 
\begin{equation}
    E_0=\lim_{n\rightarrow\infty} \min (a_1^{(K)},\ldots,a_K^{(K)}) \;
    \label{eminpds}
\end{equation}
where $a_i^{(K)}$ are the roots of Eq.(\ref{hw6}).
Remaining roots of the polynomial (\ref{hw6}) correspond to the upper bounds to the excited-state energies. 

\subsection{Variational Quantum Eigensolver}

Due to the severe limitations imposed to NISQ devices, procedures that can, in principle, determine the exact ground state energy, e.g., phase estimation, are not feasible. Instead, the variational quantum eigensolver builds on the variational principle to approximate the ground state via a parameterized circuit ansatz.\cite{vqe} These parameters are thus varied in order to minimize the expectation value of the Hamiltonian:

\begin{equation}
    \label{eq:rayleigh}
    E \leq \min_{\Psi}\langle \Psi | \hat{H} | \Psi \rangle = \min_{\vec{\theta}}\langle \Psi(\vec{\theta}) | \hat{H} |\Psi(\vec{\theta}) \rangle 
\end{equation} 
where $\vec{\theta}$ is the set of variational parameters.
Within this framework, the quantum computer is in charge of those tasks which prove to be classically challenging, i.e., to prepare the trial states $|\Psi(\vec{\theta})\rangle$ and subsequently carry out the measurements that enable the estimate of the expectation values of each term of the Hamiltonian. 
The quantum device works in tandem with a conventional computer that performs a classical update on $\vec{\theta}$ and, constituting a typical optimization cycle that goes on until the energy is found below a user-defined threshold.

The focal point in electronic structure theory for the past few decades has been quantification and characterization of electron correlation, which in quantum computing jargon is equivalent to accounting for the entanglement between electrons. Starting from a mean-field approximation $|\phi\rangle$, typically Hartree--Fock, which can be efficiently determined classically, one is then left with assigning a functional form for the ansatz $\Psi(\vec{\theta})$. Contrary to the ordinary coupled cluster (CC) theory, which abdicates the use of a unitary ansatz due to the lack of truncation in the underlying Baker--Campbell--Hausdorff expansion,\cite{shavitt_bartlett_2009} the unitary version of CC\cite{ucc1, ucc2, ucc3} is amenable to implementation in quantum hardware as it favors the adoption of unitary operators. Thus, a natural ansatz choice is:

\begin{equation}
    \label{eq:ucc}
    |\Psi(\vec{\theta})\rangle = \hat{U}(\vec{\theta})|\phi\rangle =  \text{exp}(\sum_k\theta_k\hat{\tau}_k)|\phi\rangle
\end{equation}
where $\hat{\tau} = \hat{T}_k - \hat{T}^\dagger_k$ furnishes anti-Hermitian operators from the many-body excitation operators $\hat{T}_k$ by construction. Straightforward application of the operators in Equation \ref{eq:ucc} would require up to $N$-qubit gates, with $N$ being the number of spin-orbitals, which is not prone to implementation in logical quantum devices. This can be circumvented by ``Trotterizing'' the ansatz, akin to the Trotterization in time evolution, which delivers its so-called disentangled form\cite{ucc_exact} and can be better accommodated by quantum hardware:

\begin{equation}
    \text{exp}(\sum_k\theta_k\hat{\tau}_k)|\phi\rangle \approx \prod_k\text{exp}(\theta_k\hat{\tau}_k)|\phi\rangle
\end{equation}
which is in turn no longer invariant with respect to the order in which the operators $\hat{\tau}_k$ are introduced and can lead to significant deviations in the resulting state/energy.\cite{trotter_uccsd}

It is imperative to keep in mind that electronic structure theory revolves around fermions, whose operator algebra is different from that which is naturally implemented in quantum hardware. This can be addressed by first recognizing that it is convenient to represent the Hamiltonian in second quantization. The occupation of spin-orbitals can be readily associated with the two-qubit states, such that these second-quantized fermionic operators can be mapped onto spin counterparts by a transformation that incorporates the proper antisymmetry of fermionic modes. For this purpose, the Jordan--Wigner mapping is used throughout this paper.\cite{JW} The energy expression in Equation \ref{eq:rayleigh} can be recast into a form that is more in line with the outcome of the operation of a quantum processor:

\begin{equation}
    \label{eq:H}
    E \leq \min_{\vec{\theta}} \sum_k \langle \Psi(\vec{\theta}) | \alpha_k \hat{h}_k |\Psi(\vec{\theta}) \rangle  = \min_{\vec{\theta}}\sum_k \alpha_k\text{Tr}(\hat{\rho}(\vec{\theta})\hat{h}_k)
\end{equation}
with $\hat{H} = \sum_k \alpha_k \hat{h}_k$, where each $\hat{h}_k$ is a Pauli word, $\alpha_k$ is a scalar, and $\hat{\rho}(\vec{\theta}) = | \Psi(\vec{\theta}) \rangle \langle \Psi(\vec{\theta})|$ is the density matrix.

\subsection{State Preparation Ansatz}
\label{ssec:ansatz}

Here we discuss the relevant considerations in each of the ansatz choices that we employ in this work. We adopt the convention that the qubit registry is initialized with all qubits at the $|0\rangle$ state. Upon mapping the spin-orbitals defining the electronic structure of a molecule onto these qubits, we see that the initial state of the qubit registry corresponds to the physical vacuum, that is, no orbital is occupied. Conversely, a qubit in the $|1\rangle$ state represents an occupied spin-orbital. Given that we are dealing with molecular systems, the most common choice for the quantum state would be Hartree--Fock (HF). In this case, once spin-orbitals are mapped onto qubits, bit-flip (\texttt{X}) gates are applied to those qubits identified with occupied spin-orbitals. The absence of two-qubit gates signifies that there is no entanglement between qubits, which is in agreement with the underlying premise of HF being a one-particle model, lacking electron correlation.

In order to improve upon the prepared state so we can observe improved convergence of the underlying moments expansion, but still be able to cope with the complexity of the corresponding circuit, keeping it at a manageable depth, we can test the performance of ans\"{a}tze comprised of a single many-body rotation away from the HF state, that is, containing only a single highly important many-body operator. This can be easily accomplished by the Adaptive  Derivative-Assembled Pseudo-Trotter algorithm coupled with VQE, that is, ADAPT-VQE,\cite{adapt, tang2019qubit} which we will refer to as AVQE for short. Further details of the algorithm can be found elsewhere, and we limit ourselves to the ingredients of the algorithm that are relevant to the current context. One determines a set of operators that are believed to provide the necessary physics to be able to rotate the initial state, in our case restricted HF, into the true ground state via the sequential action of these operators. We stress that starting from HF is not a requirement, but can be seen as a simple and sensible choice in molecular simulations, and many other possibilities may present themselves. Subsequently, the commutators of each of these operators with the Hamiltonian are measured, due to evaluating the derivative with respect to the variational parameters related to each operator, and the algorithm proceeds by selecting the operator with the steepest direction in the energy landscape with respect to the origin, where the HF state is located. At this point, the associated variational parameter is determined through a conventional classical optimization routine, along with any other previously selected operator. Practically, this last step is delegated to a usual VQE call.

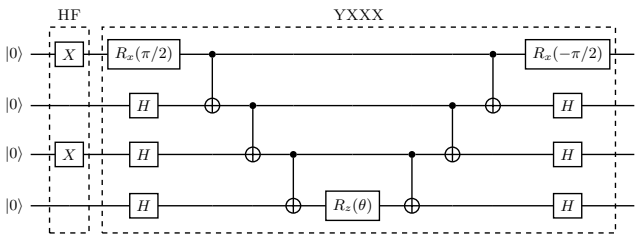
\begin{figure}

\resizebox{\columnwidth}{!}{
    \centering

 \begin{quantikz}
\lstick{$\ket{0}$} & \gate{X}\gategroup[4,steps=1,style={dashed, inner xsep=0pt},background]{{\sc HF}} & \gate{R_x(\pi/2)}\gategroup[4,steps=9,style={dashed, inner xsep=0pt},background]{{\sc YXXX}} & \ctrl{1} & \qw & \qw & \qw & \qw & \qw & \ctrl{1} & \gate{R_x(-\pi/2)} & \qw\\
\lstick{$\ket{0}$} & \qw & \gate{H} & \targ{} & \ctrl{1} & \qw & \qw & \qw & \ctrl{1} & \targ{} & \gate{H} & \qw \\
\lstick{$\ket{0}$} & \gate{X} & \gate{H} & \qw & \targ{} & \ctrl{1} & \qw  & \ctrl{1} & \targ{} & \qw & \gate{H} & \qw\\
\lstick{$\ket{0}$} & \qw & \gate{H} & \qw & \qw & \targ{} & \gate{R_z(\theta)} & \targ{} & \qw & \qw & \gate{H} & \qw
 \end{quantikz}
 }
    \caption{\label{fig:circuit}Sample circuit showing the HF state preparation and gates necessary to prepare the state upon action of a generic Pauli word of the type $Y_iX_jX_kX_l$. The f-AVQE circuit would be in the order of 8-12x as deep, as the \texttt{YXXX} portion of the circuit above is repeated roughly 8-12x times, with the addition of some single-qubit gates in between.}
    
\end{figure}

In this work we employ two distinct sets of operators in conjunction with the ADAPT-VQE algorithm. One is the fermionic singlet-adapted single and double excitations and the other one is comprised of all Pauli words of the form $Y_iX_j$ and $Y_iX_jX_kX_l$ with $\{i, j, j, l \in \mathbb{R} \wedge 0 \leq i, j, k, l \leq N-1 \}$ and $N$ being the number of qubits/spin-orbitals.\cite{mccaskey2019quantum} These two possibilities will be labeled as f-AVQE and p-AVQE for the remainder of this paper unless stated otherwise. While the former enforces $\langle \hat{S}^2\rangle=0$, it leads to much deeper circuits than the latter, as alluded to by Figure \ref{fig:circuit}. Moreover, the underlying reference is not always best described as a singlet, in which case the latter may grant more flexibility to the ansatz.

\subsection{DUCC Hamiltonian}
\label{ssec:ducc}

The double unitary coupled cluster (DUCC) formalism is predicated on the explicit decoupling of excitations describing correlation effects in and out of an active space (see Refs. \onlinecite{bauman2019downfolding,kowalski2020sub}), i.e.,
\begin{equation}
\ket{\Psi_{\rm DUCC}}=e^{\hat{\sigma}_{\rm ext}} e^{\hat{\sigma}_{\rm int}}\ket{\phi} \;,
\end{equation}
where $\sigma_{\rm int}$ and $\sigma_{\rm ext}$ are anti-Hermitian cluster operators 
\begin{equation}
\hat{\sigma}_{\rm int}^{\dagger}=-\hat{\sigma}_{\rm int} \;, \;\;\;
\hat{\sigma}_{\rm ext}^{\dagger}=-\hat{\sigma}_{\rm ext} \;.
\end{equation} 
The exact 
$\hat{\sigma}_{\rm int}$ and $\hat{\sigma}_{\rm ext}$ operators can effectively be approximated using the UCC formalism, i.e., 
\begin{equation}
\hat{\sigma}_{\rm int} \simeq \hat{T}_{\rm int}-\hat{T}^{\dagger}_{\rm int} \;, \;\;\;
\hat{\sigma}_{\rm ext} \simeq \hat{T}_{\rm ext}-\hat{T}^{\dagger}_{\rm ext} \label{exts} \;,
\end{equation}
where $\hat{T}_{\rm int}$ and $\hat{T}_{\rm ext}$ are CC-like cluster operators
producing excited configurations within and outside of the active space when acting on the reference function $\ket{\phi}$.

The foundation for the DUCC downfolding formalism is that the energy $E$, can be obtained by diagonalizing the effective Hamiltonian
$\bar{H}_{\rm ext}^{\rm eff(DUCC)}$ in the corresponding active space (defined by the projection operator $\hat{P}+\hat{Q}_{\rm int}$, where $\hat{P}$ and $\hat{Q}_{\rm int}$ are projectors onto the reference function and  orthogonal determinants in the active space, respectively) 
\begin{equation}
        \bar{H}_{\rm ext}^{\rm eff(DUCC)} e^{\hat{\sigma}_{\rm int}} \ket{\phi} = E e^{\hat{\sigma}_{\rm int}}\ket{\phi}
\end{equation}
where
\begin{equation}
        \bar{H}_{\rm ext}^{\rm eff(DUCC)} = (\hat{P}+\hat{Q}_{\rm int}) \bar{H}_{\rm ext}^{\rm DUCC} (\hat{P}+\hat{Q}_{\rm int})
\end{equation}
and
\begin{equation}
        \bar{H}_{\rm ext}^{\rm DUCC} =e^{-\hat{\sigma}_{\rm ext}}H e^{\hat{\sigma}_{\rm ext}} \;.
\end{equation}
One constructs the approximate many-body form of $\bar{H}_{\rm ext}^{\rm eff(DUCC)}$ using approximate form of 
$\hat{\sigma}_{\rm ext}$ ($\hat{T}_{\rm ext}$) based on arguments from perturbative techniques. 

The process of approximating $\bar{H}_{\rm ext}^{\rm eff(DUCC)}$ involves three critical issues: (1) length of the commutator expansion for $e^{-\hat{\sigma}_{\rm ext}}H e^{\hat{\sigma}_{\rm ext}}$, (2) rank of many-body effects included in $\bar{H}_{\rm ext}^{\rm eff(DUCC)}$, 
and (3) approximate representation of $\hat{T}_{\rm ext}$. In this paper, we follow the approximate $\bar{H}_{\rm ext}^{\rm eff(DUCC)}$ construct introduced in Ref.~\onlinecite{bauman2019downfolding} that is 
consistent through second-order, 
the $\hat{T}_{\rm ext}$ operator is expressed in terms of the external part of the standard CCSD cluster operator, and one- and two-body terms are included in $\bar{H}_{\rm ext}^{\rm eff(DUCC)}$.
The only difference being that we employ the full form anti-symmetrized dressed two-electron integrals, in contrast to Refs.~\onlinecite{bauman2019downfolding} and \onlinecite{bauman2019quantum}, where it was assumed that Mulliken orbital-type  dressed integrals $({\bf P} {\bf Q}|\bf{R} \bf{S})$ are
obtained from the $({\bf P}(\alpha) {\bf Q}(\beta)|\bf{R}(\alpha) \bf{S}(\beta))$ subset. We refer the reader to Refs.~\onlinecite{bauman2019downfolding} and \onlinecite{bauman2019quantum} for full details about the DUCC ansatz.

\section{Results}

The standard and DUCC second-quantized Hamiltonians used here are obtained from the NWChem software,\cite{NWChemPPF} and subsequently mapped onto the corresponding spin Hamiltonians via the Jordan--Wigner transformation. All quantum simulations are carried out with the XACC quantum-classical framework,\cite{xacc1, xacc2} with the QPP\cite{qpp} accelerator as well as the tensor network quantum virtual machine (TNQVM)\cite{tnqvm} together with the ITensor visitor as the virtual backend.\cite{itensor} In the current work, the energies are computed as expectation values over the proper qubits and in the absence of noise, akin to the simulation of the underlying state vector subject to infinite sampling.

The quantum connected moments expansion algorithm proposed recently\cite{kowalski2020cmx} is mainly focused on utilizing the quantum resources to compute the standard moments $K_n=\langle\phi | H^n | \phi \rangle$, which then on the classical side constitute the connected moments and their expansion formulation to reconstruct the energy of the system.\cite{horn1984t,cioslowski1987connected,peeters1984upper, soldatov1995generalized} As briefly reviewed in Ref. \citenum{kowalski2020cmx} there exist several re-summation techniques utilizing the (connected/canonical) moments to reconstruct the energy including the original $t$-expansion based on Horn--Weinstein theorem,\cite{horn1984t} Cioslowski's recursive expansion,\cite{cioslowski1987connected} Knowles's generalized Pad\'{e} approximation,\cite{knowles1987validity} alternate moments expansion,\cite{mancini1994analytic} generalized moments expansion,\cite{fessatidis2006generalized} and Peeters--Devereese--Soldatov variational formulation.\cite{peeters1984upper, soldatov1995generalized} The present work will focus primarily on the PDS formulation based on the quantum evaluation of the standard moments, and CMX to a smaller extent, which are coupled to an ADAPT-VQE-optimized form of the trial wave function and used to construct the PDS($K$)/CMX($K$) energy.

\begin{figure}
    \centering
    \includegraphics[width=\columnwidth]{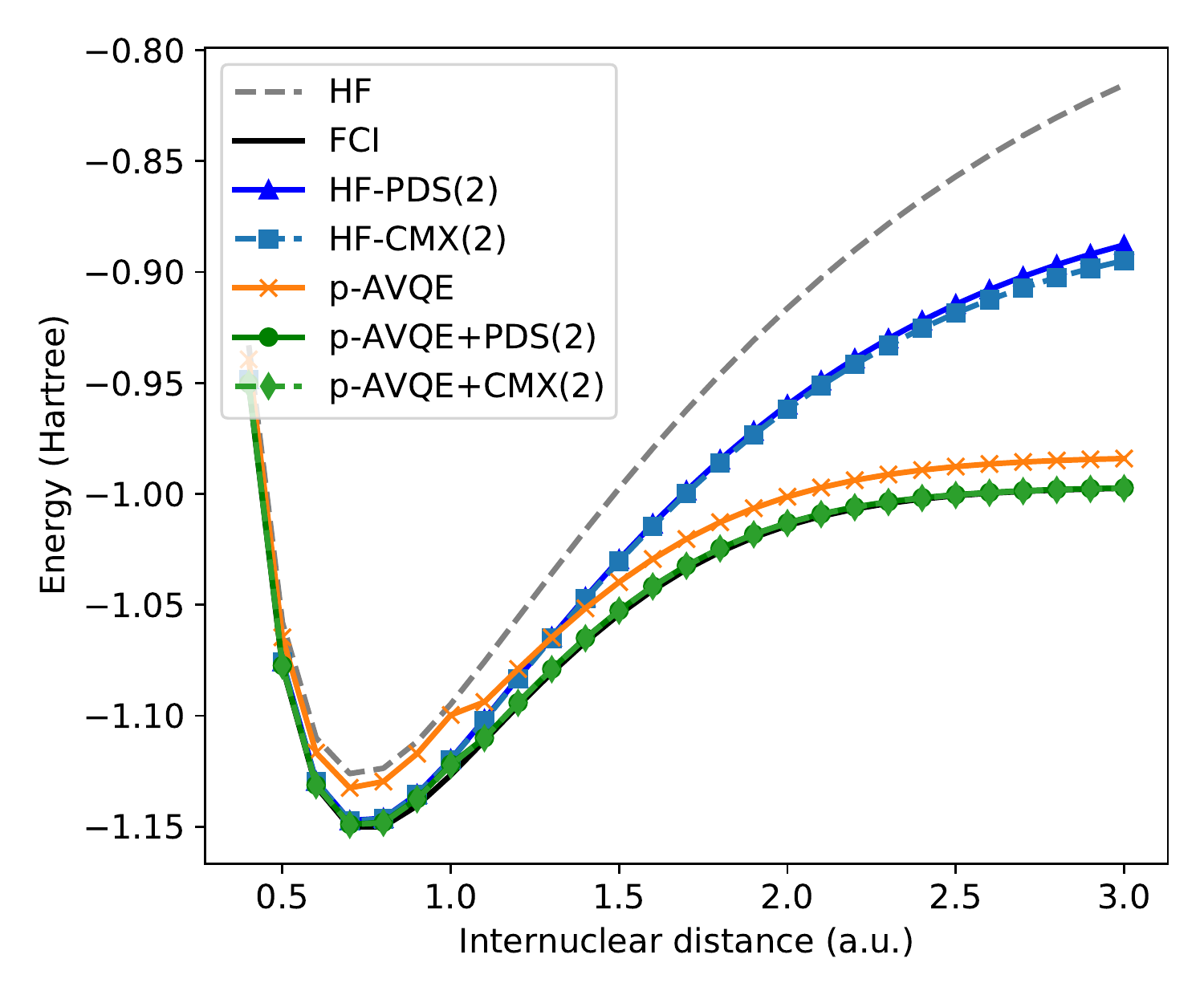}
    \caption{\label{fig:h2_pec}Potential energy curves of H$_2$ in the 6-31G basis set computed with Hartree--Fock (HF), full configuration interaction (FCI), PDS(2) with the HF state (HF-PDS(2)), CMX(2) with the HF state (HF-CMX(2)), single iteration ADAPT-VQE with the Pauli operators (p-AVQE), PDS(2) with the Pauli-VQE state (p-AVQE+PDS(2)), CMX(2) with the Pauli-VQE state (p-AVQE+CMX(2)).}
\end{figure}

We now turn to the hydrogen molecule in the 6-31G\cite{631g} basis set and investigate the potential energy curve (PEC) obtained upon the different choices of ansatz and their effect onto the PDS(2) and CMX(2) expansions. The HF in the restricted formalism is the mean-field treatment in the entire PEC, and because the underlying reference is a singlet and the operators in the f-AVQE simulations are linear combinations of elements in the Pauli operator pool, the energy is the same for these in the two cases, so we report solely the p-AVQE results in Figure \ref{fig:h2_pec}.

We first observe the expected behavior from HF, that is, that it cannot properly undergo a homolytic bond dissociation, thus its inability to even qualitatively track the FCI energies. There is significant correlation energy recovery by turning to either PDS(2) or CMX(2) close to the equilibrium region of the potential energy curve. This is also to be expected since this is the region where the FCI expansion is largely dominated by the HF configuration, and as the HF contribution to the FCI wave function decreases, so does the overlap between the two, which implies that either choice of CMX is also negatively impacted. When looking at the p-AVQE results, we start by pointing out that it displays remarkably different accuracy depending on the location in the curve. It closely resembles the HF curve in the equilibrium region, while following the qualitative trend of FCI as the H$_2$ is stretched. There is a clear discontinuity in this curve, resulting from the fact that these two distinct regimes are addressed by the ADAPT-VQE algorithm selecting a different operator in each. When the state proposed by the p-AVQE algorithm is used in conjuntion with PDS(2)/CMX(2), an interesting behavior emerges. The p-AVQE+PDS(2) energies are now found practically on top of the corresponding FCI numbers as well as the those obtained with the p-AVQE+CMX(2) scheme. This is quite remarkable given that the p-AVQE curve shows clear qualitative and quantitative shortcomings as the ADAPT-VQE algorithm is far from converged, yet the resulting state builds enough overlap with the exact ground state that even a low-order CMX, in this case PDS(2) or CMX(2), is able to reach FCI-level accuracy.

Comparing CMX(2) and PDS(2), there is little difference between the two from visual inspection of Figure \ref{fig:h2_pec}. Overall, CMX(2) slightly outperforms PDS(2), and such a deviation becomes more pronounced as the bond is stretched. Despite this allegedly superiority, PDS(2) energies are variational, while we do observe CMX(2) energies falling below the associated FCI energies. Additionally, PDS($K$) provides $K-1$ energy values, approximating the electronic energy spectrum of the corresponding Hamiltonian, whereas CMX, regardless of the order, is limited to approximating the ground state alone. As will also become clear later in this article, PDS reduces the error from FCI at a more rapid rate as the expansion order is increased than CMX. Thus, we proceed by prescribing the PDS expansion and employ it for most of the remainder of this paper.

\begin{figure}
    \centering
    \includegraphics[width=\columnwidth]{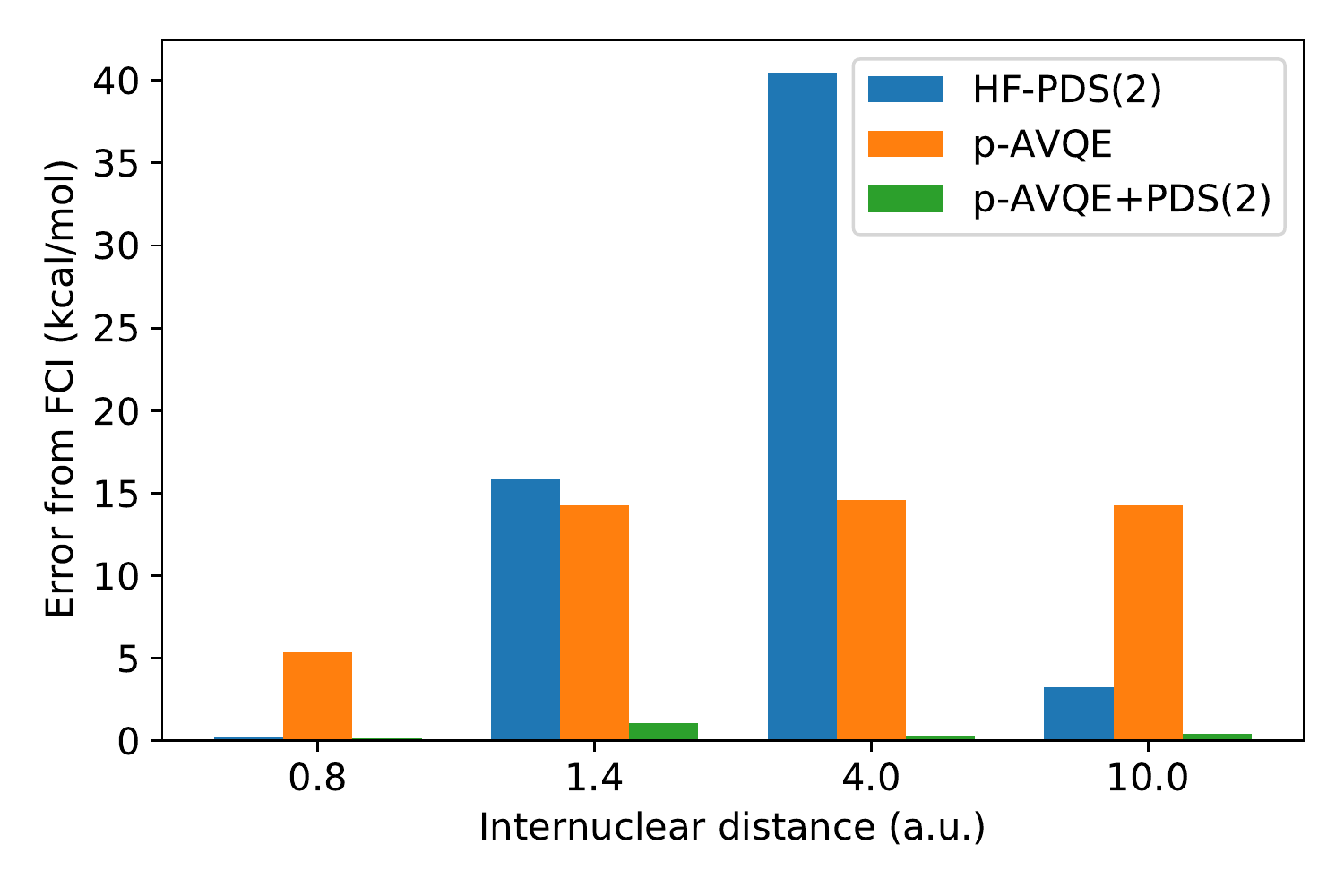}
    \caption{\label{fig:h2_ducc}Errors with respect to FCI for H$_2$ within the cc-pVTZ basis set downfolded to an active space of four spatial orbitals with the DUCC Hamiltonian for PDS(2) with the HF state (HF-PDS(2)), single iteration ADAPT-VQE with the Pauli operators (p-AVQE), and PDS(2) with the aforementioned p-AVQE state (p-AVQE+PDS(2)).}
\end{figure}

Turning to a different basis set, cc-pVTZ\cite{dunning} is employed to construct a DUCC Hamiltonian for the H$_2$ molecule where the virtual orbital space is downfolded onto an active space of four spatial orbitals (one occupied and three virtual). The results for simulations with this Hamiltonian in four distinct bond lengths are provided in Figure \ref{fig:h2_ducc}.

The errors in HF energies are orders of magnitude larger than the other methods being studied, so we do not plot them in Figure \ref{fig:h2_ducc}. PDS(2) is only capable of achieving chemical accuracy for bonds shorter than the equilibrium distance (0.8 a.u.), and these energies deteriorate for longer bonds. The ansatz formed with the operator selected by p-AVQE is not able to achieve quantitative agreement with the reference FCI energies in any of the geometries. Yet, here again this state can drastically reduce the errors in PDS(2) energies, which are now all found below the 1 kcal/mol chemical accuracy mark, except for the 1.4 a.u. bond length, which is borderline chemically accurate (1.069 kcal/mol).

\begin{figure}
    \centering
    \includegraphics[width=\columnwidth]{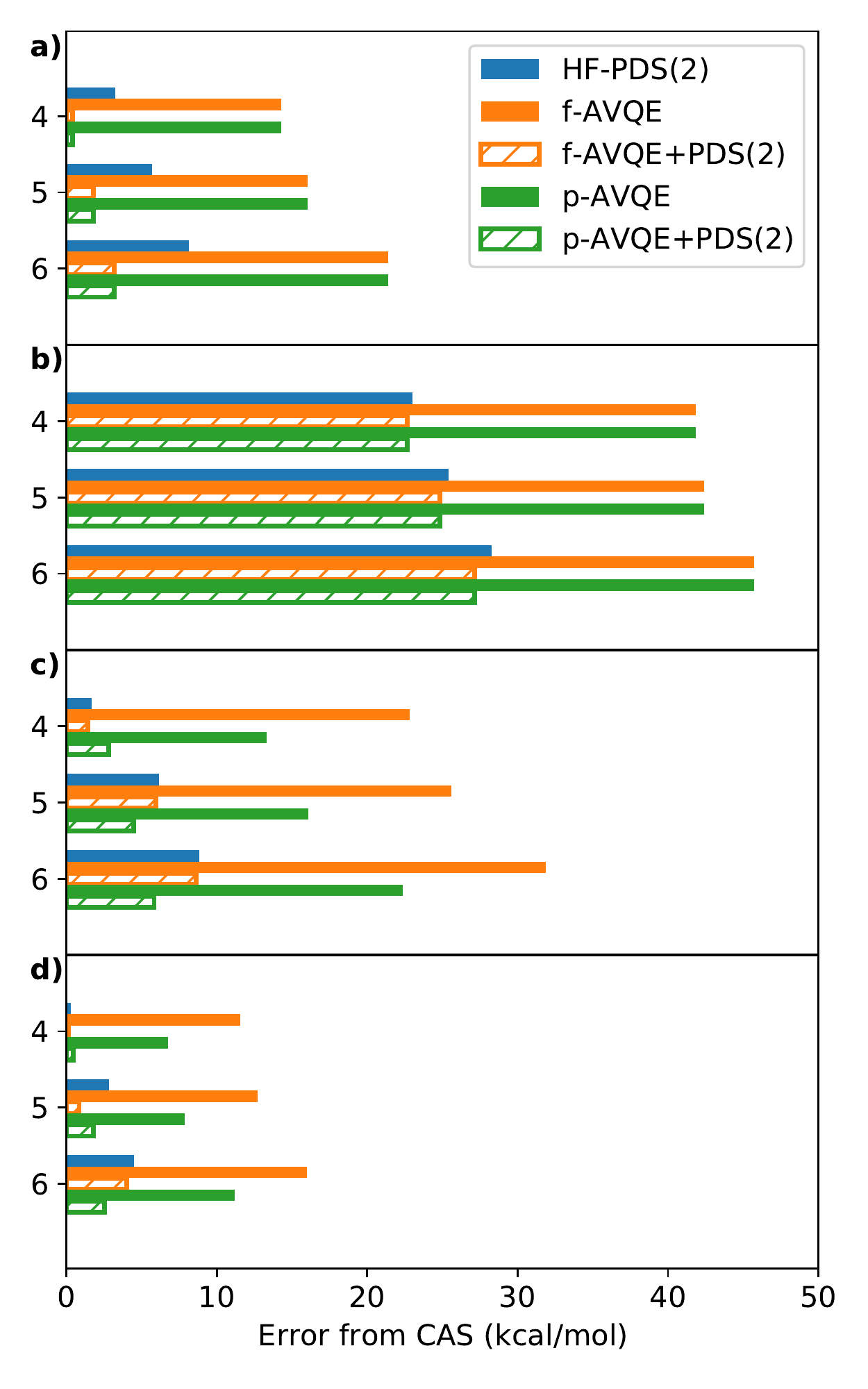}
    \caption{\label{fig:h4_321g}Errors with respect to the CAS energies of the lowest energy singlet and triplet states in Hartree--Fock (HF), PDS(2) with the HF state (HF-PDS(2)), single iteration ADAPT-VQE with the Pauli operators (p-AVQE), and PDS(2) with the aforementioned p-AVQE state (p-AVQE+PDS(2)) in the different active spaces (y-axis) for the H$_4$ molecule in the 3-21G basis set with adjacent hydrogens 2 {\AA} apart in a) singlet state, linear geometry; b) singlet state, square geometry; c) triplet state, linear geometry; d) triplet state, square geometry.}
\end{figure}

A considerably more challenging scenario is the H$_4$ molecule in two distinct spatial configurations: linear and planar square. Simulations of H$_4$ in these two geometries with the 3-21G\cite{321g} basis where the distance between adjacent hydrogens is 2 {\AA} are presented in Figure \ref{fig:h4_321g}, where we report the deviations from the reference CAS energies with three different active space sizes, aiming at the lowest energy singlet and triplet states.

Before discussing specific details in these plots, let us turn our attention to some general trends. The linear geometry is best represented by a singlet state, while the planar one is mostly associated with a triplet state, and this is further corroborated by the scale of the corresponding plots. Given that the ans\"{a}tze generated by ADAPT-VQE is comprised of a single operator, with the circuit depth being roughly constant for a certain choice of operator pool, the increase in the size of the active space is expectedly accompanied by an increase in error with respect to the associated CAS energies, as the corresponding states would be farther from the ground state when active spaces are larger.

Once more, it is apparent that there is no numerical discrepancies between the results with the f-AVQE and p-AVQE for singlet states (top two plots). The PDS(2) energy on top of the HF state alone represents a substantial improvement over the energy obtained for the more complicated ADAPT-VQE circuits for the linear H$_4$ in a singlet state. Albeit the energy computed from these ans\"{a}tze not presenting an advantage over plain PDS(2), the respective prepared states grant further improvement in the PDS(2) computation. Changing the geometry, we see that the results with the planar configuration fall into two main categories, namely energies based in PDS(2) all flock around the same value, and the same is true for the results with ADAPT-VQE. The PDS(2) results are significantly closer to the corresponding CAS numbers than ADAPT-VQE, while the state from the latter does not provide any appreciable improvement over PDS(2) obtained with the HF state. This is in agreement with the understanding that the linear H$_4$ is best represented by a singlet state, whereas the planar H$_4$ geometry is most in line with a spin triplet.

The fact that the operators in the f-AVQE simulations are linear combinations of the operators in the operator pool of Pauli operators explains why there is no difference between these two instances for singlet states. Moving to the same molecules treated under a triplet state, we see this is no longer the case. Given that the Pauli operators are not constrained to a specific spin state, p-AVQE can achieve better energetics than f-AVQE. The PDS(2) energies for all state preparation possibilities being considered here building on a triplet reference show an overall significant move toward the reference CAS energies, as reported in Figure \ref{fig:h4_321g}c-d. 

\begin{figure}
    \centering
    \includegraphics[width=\columnwidth]{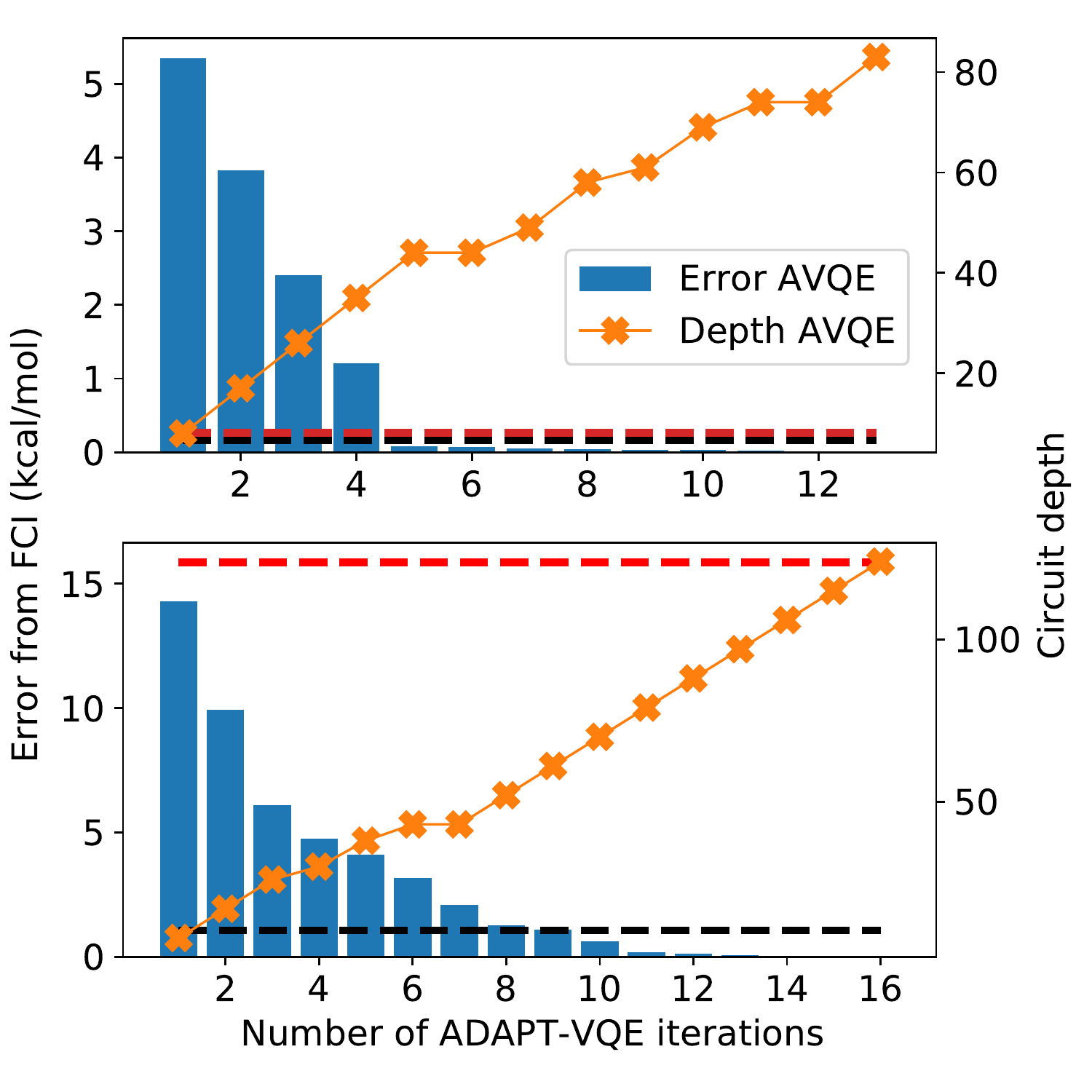}
    \caption{\label{fig:h2_adapt}Convergence of the ADAPT-VQE energy and the corresponding circuit depth with the operator pools of Pauli operators for the H$_2$ molecule with DUCC Hamiltonians in which the cc-pVTZ orbital space is downfolded to 4 spatial orbitals at 1.4 (top) and 4.0 a.u. (bottom). The errors in the PDS(2) and p-AVQE+PDS(2) energies are represented by the red and black dashed lines and refer to PDS(2) energies using the HF state and the state prepared upon the first iteration of ADAPT-VQE (p-AVQE). The circuit depth for HF-PDS(2) is 1 and the circuit depth for p-AVQE+PDS(2) is 8.}
\end{figure}

In order to develop some intuition regarding how PDS(2) improves on the application of VQE-type strategies alone, here illustrated by ADAPT-VQE, and can provide a much more efficient alternative energy functional when taking into account considerations of implementation and execution on actual quantum hardware, we revisit two of the H$_2$ DUCC Hamiltonians for the H-H length of 1.4 and 4.0 a.u. as the former is representative of a weak correlated system while the latter is found in a more strongly correlated regime. In Figure \ref{fig:h2_adapt}, we plot the convergence of p-AVQE until the gradient norm falls below the 10$^{-2}$,\cite{adapt} while each individual VQE optimization is converged with the aid of parameter shift gradients\cite{parameter_shift} to a threshold of 10$^{-7}$ Hartree in absolute energies.

Because the ground state at 1.4 a.u. is largely dominated by HF, the state prepared by ADAPT-VQE (p-AVQE) does not provide a major improvement in the energies versus the HF-PDS(2) energy, so it can be concluded that it is the PDS(2) procedure itself that is the main responsible for the correlation energy recovery in the top plot of Figure \ref{fig:h2_adapt}. ADAPT-VQE provides a means to grow an ansatz in an incremental and somewhat controlled manner, and we infer that it needs at least 5 iterations with the Pauli operators for the expectation value of the Hamiltonian to surpass the PDS(2) energy estimate. On the other hand, due to the pronounced stretching of the bond at 4.0 a.u., the restricted HF reference is quite inadequate, and it is corroborated by the HF-PDS(2) energy being less satisfactory than the ADAPT-VQE energy at the first iteration, which we see that at this point is still far from converged. Yet, here again, the interplay between PDS(2) and even the simplest ADAPT-VQE state provides a remarkable enhancement, which is only matched by ADAPT-VQE alone after 9 iterations. Needless to say, in both cases the circuit proposed by ADAPT-VQE is much deeper than that involved in the p-AVQE+PDS(2) for the former to achieve comparable accuracy with the latter, in accordance with the bottom plots in Figure \ref{fig:h2_adapt}.\cite{adapt_benchmark} On that note, it is also noteworthy that, while PDS(2) requires computations of $\langle H^3 \rangle$, which can involve many terms, many more than any individual expectation value found in ADAPT-VQE, it is a "one-shot" calculation, where all measurements are employed in the energy computation. On the other hand, ADAPT-VQE needs to recompute the commutators of every single operator in the pool at each iteration, incurring in a much larger number of measurements if the algorithm needs a significant number of operators to converge, but which are not further utilized throughout the algorithm (but could in principle be). This is further aggravated by the fact that these measurements are carried out with an increasingly deep circuit during the ADAPT-VQE procedure.

\begin{figure}
    \centering
    \includegraphics[width=\columnwidth]{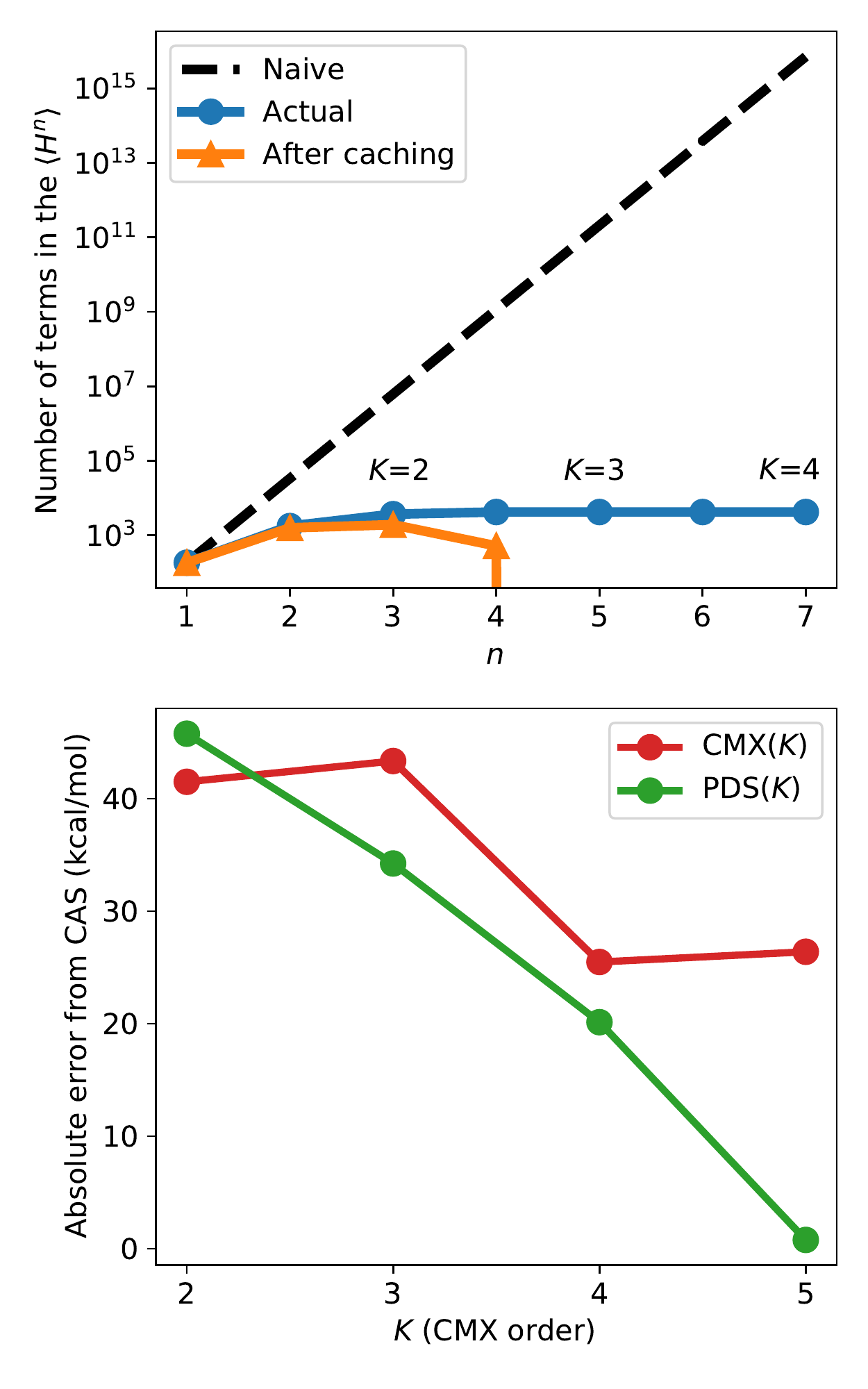}
    \caption{\label{fig:measurements}(top) Number of unitaries to be measured for Hamiltonian powers, $H^n$ ($n=1-7$), for the square H$_4$ in the STO-6G basis before and after Pauli reduction, labeled by Naive and Actual, respectively, and also the necessary number of measurements following caching. Note that after $n=4$ the number of terms to be measured converges to a constant which covers all the necessary unitaries, when combined with different set of scalar coefficients, that constitute arbitrary $H^n$. The CMX order that can be computed up to a given $n$ is shown by the value of $K$ ($n=2K-1$). (bottom) Error from the reference CAS energy for the different expansion orders with the measured terms from the top plot.}
\end{figure}

Intuitively, one could (naively) expect that an increase in the order of the desired CMX is accompanied by exponentially more measurements, which would, in turn, diminish the appeal of the CMX-based schemes as a viable implementation target on NISQ hardware. As it turns out, the Hamiltonian shows a tendency to a natural truncation such that the number of necessary circuit implementations eventually plateau, regardless of the CMX order. This has two important implications. The first is that after this plateau is achieved, an arbitrary order CMX energy can be obtained by simple classical post-processing, with no further need to resort to the quantum co-processor. Second, a caching mechanism can be put in place whereby only circuits corresponding to Pauli words not been previously measured need to be implemented. This is illustrated by Figure \ref{fig:measurements} taking the square planar H$_4$ in the STO-6G basis set.

Even though we here explore more frugal alternatives, an important result to point out that if the level of accuracy that they can achieve is not satisfactory, chances are they can be improved by increasing the CMX order without an insurmountable cost. The moments of the Hamiltonian naturally reach a truncation in the number of terms, and with caching, an arbitrary order CMX energy can be computed with a fixed number of measurements. In this case, $\approx 4000$ terms would be sampled, after which any CMX order would be accessible at classical post-processing cost. From the top plot in Figure \ref{fig:measurements} it is concluded that no new term needs to be measured after $\langle H^4 \rangle$, which is part of PDS(3)/CMX(3). Therefore the 4th and 5th order PDS/CMX energies reported in the bottom plot are ``free'' as far as quantum resources are concerned, but lead to massive energetic improvement.

\begin{figure}
    \centering
    \includegraphics[width=\columnwidth]{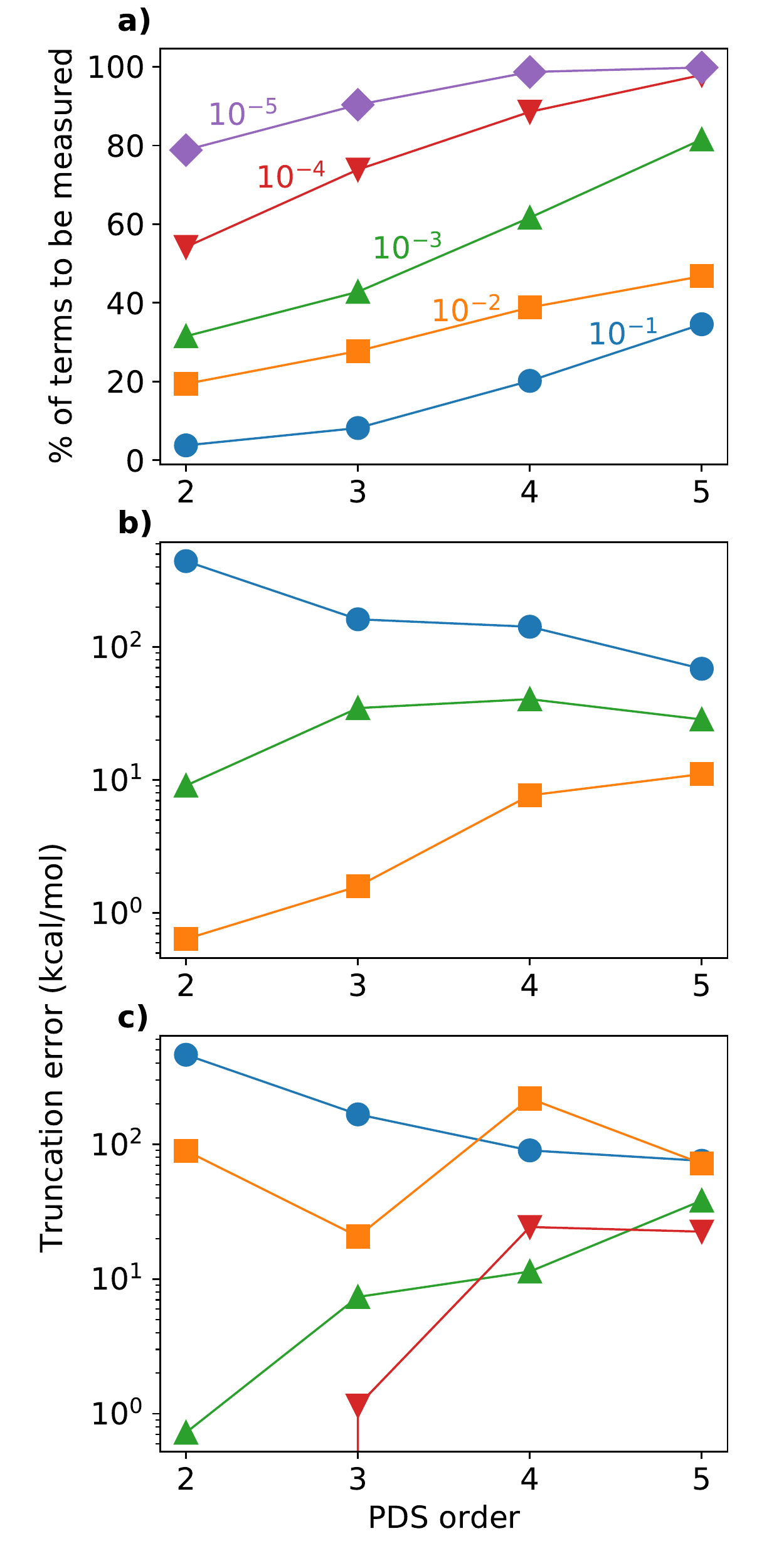}
    \caption{\label{fig:threshold}a) Fraction of Hamiltonian terms that need to be measured; b) corresponding truncation error in HF-PDS and c) p-AVQE+PDS or different choices of threshold as a function of PDS order in the square H$_4$ molecule. The center and bottom plots are in logarithmic scale, so missing points/lines represent zero truncation error.}
\end{figure}

It should be noted that the caching strategy introduced above is not an approximation and is possible owing to all measurements being taken from the same state preparation circuit. Upon the appropriate rotations to the Z-basis, the measurement step is akin to the projection onto the Z-axis of a given qubit, thus is constrained to $\leq |1|$, which is in turn modulated by the scalars $\alpha_k$ (Eq. \ref{eq:H}). Caching can therefore be made more efficient by bypassing the measurement of terms deemed of negligible importance, i.e., terms of a given moment whose scalar coefficient is found below a chosen threshold are ignored. The impact of different choices for this threshold can be seen in Figure \ref{fig:threshold}, still taking the square H$_4$ molecule as a case study.

The top plot in Figure \ref{fig:threshold} shows that the percentage of terms in the moments to compute PDS to a certain order whose coefficient is above the set threshold increases as the the threshold is made smaller, which is to be expected. This fraction also increases monotonically with the CMX order, which is invariant with respect to the specific expansion flavor, and we limit ourselves to PDS here. The reader is reminded here that PDS is variational, and in analyzing the center and bottom plots in Figure \ref{fig:threshold} we find that PDS($K$) results do not necessarily improve with increasing $K$ for loose values of the threshold, which signals that, by virtue of missing too many of their terms, the remaining pieces of the connected moments refrain the PDS expansion from displaying its variational property. We refer to truncation errors as the deviations due to the removal of terms with coefficients smaller than the threshold, and observe that such errors can grow with increasing PDS order, which is the case when the threshold is set to 10$^{-3}$, and to a smaller extent, 10$^{-2}$ for HF-PDS, effect which is more pronounced for the p-AVQE state. 

It is somewhat surprising and counter-intuitive that a tighter threshold leads to larger errors, but as we go to higher moments (Eq. \ref{hw3}), the coefficients of the surviving terms do not rise trivially, as observed with for the entirety of the 10$^{-2}$ and 10$^{-3}$ plots in Figure \ref{fig:threshold}b and for the PDS(4) results in \ref{fig:threshold}c. For the most part, these instances yield energies that are considerably above chemical accuracy, so neither paints a reliable picture of the actual moments. Continuing the analysis of Figure \ref{fig:threshold}, the errors are plotted in logarithmic scale, thus the missing points/lines signify zero truncation errors. This means that there is no error in discarding Hamiltonian terms whose coefficients are smaller than 10$^{-4}$ when adopting the HF state. The p-AVQE in question here incurs a PDS expansion that is more sensitive to the type of truncation in consideration here, yet two important conclusions can be drawn. One is that if the goal is to remain in the low-order CMX regime, that is, PDS(2), the 10$^{-4}$ stills yields the exact, untruncated PDS(2) energy, and 10$^{-3}$ is found within chemical accuracy. The other is that 10$^{-5}$ renders exact energies regardless of the prepared state. For this relatively small Hamiltonian, it does not at first allow for as drastic savings as the other threshold. However, when taking PDS(2) for instance, it implies a reduction of 20\%, which is in the order of 800 fewer terms, each of which routinely takes $\mathcal{O}(10^3)$ shots to reach reliable statistics in realistic NISQ operation. Thus this reduction should not be underestimated, as in this relatively simple example it would mean in the order of $10^6$ circuits would not need to be implemented. We acknowledge that, while some of these conclusions may be in part specific to the problem at hand, and different thresholds may be more appropriate for different molecules, we also expect that such a compression could be further improved by turning to a Hamiltonian in a basis where more coefficients are significantly small, such as one of the many local orbitals in the literature.

\section{Conclusion}

Several quantum algorithms have been proposed for computing the standard moments. For relatively small systems, the Hadamard test can be directly applied to compute the expectation value of Hamiltonian powers.\cite{kowalski2020cmx} For a typical molecular Hamiltonian being represented by $N$ qubits, the number of measurements in the Hadamard measurement of $K^n$ scales as $\mathcal{O}(N^{4n})$, but can be reduced if Pauli strings are multiplied before the measurements and low-order moments are re-utilized as contributions to high-order moments. For larger systems, active space and local approximations can be further applied to reduce the effective system size and the number of measurements. Alternatively, $K^n$ can also be approximated by a linear combination of the time-evolution operators as proposed in recent reports.\cite{seki2020quantum,kyriienko2020quantum,bespalova2020hamiltonian} With the impossibility of reliably determining the ground state of molecular systems persisting in the near future, except for much too simplistic cases, quantum connected moments expansions posit themselves as a feasible and better energy functional than the straightforward expectation value of the Hamiltonian. 

We show that such expansions greatly benefits from even simple circuits capable of building entanglement / overlap with the true ground state. For that purpose, we observe that ADAPT-VQE, upon a judicious choice of many-body operators, serves as an excellent scheme to deliver circuits that carry the prospect of being amenable to NISQ devices while incurring a significant improvement in the accuracy of the computed ground state energy estimates. In passing, we point out that there may be other options when it comes to an ansatz circuit that is constructed in a controlled manner. We have briefly investigated the so-called hardware-efficient ansatz with a single layer in order to keep the circuit depth at a minimum.\cite{kandala2017hardware} As previously noted, we also observed that the variational determination of the optimal parameters is a burdensome task, with convergence to multiple local minima from random parameter initialization, none of the energies that we found were comparable to those from HF and ADAPT-VQE circuits above. Better ground state estimates could, in principle, be found by increasing the number of layers, which goes against our principle of trying to minimize circuit depth and would also aggravate the difficult of the ansatz optimization.

Putting in place a mechanism that enables previously measured terms to be cached makes quantum CMX more attractive by substantially reducing the number of required circuit implementations, and thanks to the intrinsic structure of the Hamiltonian, only a limited number of moments' terms may need to be computed for an arbitrary CMX order, after which the strategy being proposed here implies classical resource demands exclusively. The underlying efficiency of our proposal can be considerably extended by recognizing that terms in the moments expansion whose coefficient is below $10^{-4}$ signals that they can be safely ignored, leading to further resource savings.

By laying the groundwork that points to an avenue for manageable circuit implementation and sensible and controllable number of measurements, the present contribution constitutes a coherent and viable strategy to assess the energy of molecular systems, encouraging the future deployment of CMX approaches in quantum devices, which will be presented in a forthcoming publication.

\section{Acknowledgements}

This work was supported by the “Embedding Quantum Computing into Many-body Frameworks for Strongly Correlated Molecular and Materials Systems” project, which is funded by the U.S. Department of Energy (DOE), Office of Science, Office of Basic Energy Sciences, the Division of Chemical Sciences, Geosciences, and Biosciences. 
This work was also supported by the Quantum Science Center (QSC), a National Quantum Information Science Research Center of the U.S. Department of Energy (DOE).
This research used resources of the Oak Ridge Leadership Computing Facility, which is a DOE Office of Science User Facilities supported by the Oak Ridge National Laboratory under Contract DE-AC05-00OR22725. This research used resources of the Compute and Data Environment for Science (CADES) at the Oak Ridge National Laboratory, which is supported by the Office of Science of the U.S. Department of Energy under Contract No. DE-AC05-00OR22725. This work was carried out at Oak Ridge National Laboratory, managed by UT-Battelle, LLC for the U.S. Department of Energy under contract DE-AC05-00OR22725.

\bibliographystyle{iopart-num}
\bibliography{main}
\end{document}